# A Novel Model for Optimized GSM Network Design

Alexei Barbosa de Aguiar, Plácido Rogério Pinheiro,
Álvaro de Menezes S. Neto, Ruddy P. P. Cunha, Rebecca F. Pinheiro
Graduate Program in Applied Informatics, University of Fortaleza
Av. Washington Soares 1321, Sala J-30, Fortaleza, CE, Brazil, 60811-905
alexei@verde.com.br, placido@unifor.br, netosobreira@edu.unifor.br,
ruddypaz@hotmail.com, becca.pin@gmail.com

**Abstract** – GSM networks are very expensive. The network design process requires too many decisions in a combinatorial explosion. For this reason, the larger is the network, the harder is to achieve a totally human based optimized solution. The BSC (Base Station Control) nodes have to be geographically well allocated to reduce the transmission costs. There are decisions of association between BTS and BSC those impacts in the correct dimensioning of these BSC. The choice of BSC quantity and model capable of carrying the cumulated traffic of its affiliated BTS nodes in turn reflects on the total cost. In addition, the last component of the total cost is due to transmission for linking BSC nodes to MSC. These trunks have a major significance since the number of required E1 lines is larger than BTS to BSC link. This work presents an integer programming model and a computational tool for designing GSM (Global System for Mobile Communications) networks, regarding BSS (Base Station Subsystem) with optimized cost.

## I. INTRODUCTION

The GSM mobile networks have a very sophisticated architecture composed by different kind of equipments [14].

One of the most important of these equipments, located at the core of the network, is MSC (Mobile Switching Center). MSC has many vital duties like register and unregister MS (Mobile Station), analyze call destinations, route calls, handle signaling, locate MS through paging, control handover, compress and crypt voice, etc. Indeed, it is one of the most expensive components of the network.

The HLR (Home Location Register) works as a subscriber database, storing information concerning its state, location, parameters and service data. It is constantly queried and updated by MSC.

The SGSN (Serving GPRS Support Node) are analogous to MSC but are dedicated to the packet data transmission services instead of handling voice calls. Many of its mechanics are identical or similar to its voice counterpart and deals with HLR as well.

Hierarchically below each MSC we have BSC (Base Station Controller) nodes. They are not present in IS-136 (TDMA) networks. BSC reduces the cost of the network. One of the reason is that it concentrates the processing intelligence of BTS (Base Transceiver Stations) nodes, which are the most numerous and spread equipments. Other impacting factor is that, although BSC depends on MSC for many activities, it is the first layer telephony switch, geographically concentrating traffic. This means that the trunks that carries the traffic from BSC to MSC are statistically dimensioned based on Erlang's traffic theory instead of one-by-one channel fashion.

The BTS radiates the RF (Radio Frequency) signal to the mobile phones and receive its signal back. Antennas in the top of towers or buildings radiate this RF, creating coverage areas called cells. The geographical allocation of BTS is guided by RF coverage and traffic demand.

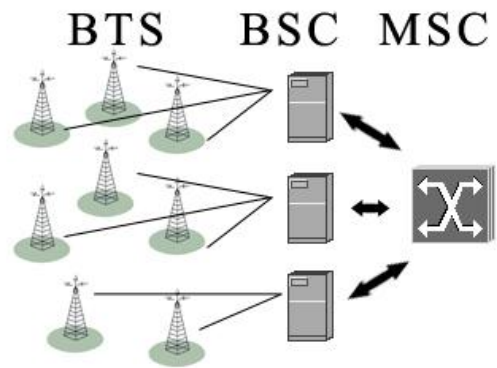

Fig.1. Mobile Network Design

The focus here will be concentrated in the BSS (Base Station Subsystem) which faces the radio resources towards MS. BSS is the group of equipments and softwares that integrates BSC nodes, BTS nodes and MSC. Transmission network plays an important role on linking them all.

The network design usually starts at the cell planning department. The coverage area required given to cell planning engineer team and the traffic is estimated by geographic regions. This region's traffic density variation can be very wide.

When coverage is the goal, RF engineers look for sites with high altitudes and free of obstacles to reach larger distances. On the other hand, when the goal is traffic, hotspots are distributed with full equipped BTS nodes. Its radio channel's power is configured lower and the RF irradiation is directed to the "near" ground with a higher antenna tilt angle.

In urban areas the BTS proximity is limited by interference since there is a limited number of RF channels and they are repeated on and on along the coverage area. The BTS sites are allocated in a triangular grid pattern, where it is possible. This allocation is due to the coverage pattern of its tree groups of antennas, disposed with 120° angles between then.

Once all BTS placements are determined with its correspondent channel dimensioning, it is possible to plan how many BSC nodes are need, witch capacity each one may have and its geographical allocation. All this factors are highly related to the choices of which BTS nodes are linked to which BSC nodes.

The links between BTS and BSC are E1 lines that hold voice channels slots. They are configured deterministically in a one-to-one basis, regarding the radio channels slots of the BTS. It is called Abis interface.

On the other hand, trunks that link BSC to MSC are E1 lines dimensioned by the total traffic from all of its BTS. It is called A interface. These trunks are similar to trunks between two MSC or other conventional telephony switches. The voice channels in these trunks are seized statistically by demand and the total number of busy channels varies during the day. All calls must pass through the MSC, even when both subscribers are very close, in the same BTS and BSC area.

The Erlang B formula calculates the blocking probability (or congestion, or Grade of Service GoS) to a given number of resources (voice channel, normally) and offered traffic.

Each one of the three variables in this formula can be calculated from the two others depending on the context. The percentile of calls that are lost can be calculated for a given number of voice channels available in some equipment and the measured traffic. To solve a congestion scenario this formula provides the number of channels that would be necessary to flow this traffic for a maximum tolerable GoS (2%, for instance). Other possibility is to calculate how much traffic can be carried with a given number of channels and the desired GoS.

The Erlang B formula eq. (1) is shown below:

$$e_b = \frac{\frac{a^n}{n!}}{\left(\sum_{i=0}^{n} \frac{a^i}{i!}\right)} \quad (1)$$

$e_b$ is the probability of blocking, also known as GoS, $n$ is the number of resources (voice channels in this case) and $a$ is the amount of traffic offered in Erlangs.

Besides channel resources, some BSC have a deterministic way of allocation for other kind of resources. When a new radio channel is installed in a BTS, some required resources (processor and memory, for instance) are associated with this new radio channel in a fixed way. These resources are compromised with the radio channel, even though it is idle. Thus, this kind of BSC has a fixed maximal capacity, for instance, 4096 radio voice channels (slots).

Some more modern BSC uses a pool of resources that are associated to radio voice channels on demand, when a call is made. This feature increases the BSC capacity. Using this type of BSC, its maximum capacity cannot be determined by its number of radio channels, but by its traffic in Erlangs. For instance, the 4096 radio voice channel BSC could be equivalent to a 4058 Erlangs (at 2% GoS) BSC model, with virtually unlimited number of radio voice channels, depending on their traffic demand.

So the A interface from BTS to BSC is made of deterministic channels in E1 lines. These lines waste transmission resources. Moreover, the A interface from BSC to MSC is made of statistical channels in E1 lines. These lines are more efficient.

It was said that BSC reduces transmission costs, but they themselves represents network design costs. It is a design tradeoff. The more BSC we distribute along the coverage area, the lower are transmission costs, since the distances between BTS to BSC decreases. On the other hand, the BSC has its acquisition cost. The balance between these two costs is reached with the optimal geographical allocation of the BSC, associated with its correct choice of model that has its respective capacity and cost.

A typical GSM network has hundred or thousand BTS and tens or hundreds of BSC. The human capacity of designing efficient networks with such magnitudes is very limited and the network costs are high. The use of computational tools can reduce these costs radically. That is what is proposed here.

## II. THE INTEGER PROGRAMMING MODEL

This is an Integer Programming model [8] capable of minimizing the total network cost and providing the design solution to achieve this minimal cost.

$T = \{t_1, t_2, t_3, \ldots, t_m\}$ BTS nodes;
$B = \{b_1, b_2, b_3, \ldots, b_n\}$ BSC nodes;
$W = \{w_1, w_2, w_3, \ldots, w_o\}$ BSC models;
$C = \{c_0, c_1, c_2, \ldots, c_p\}$ Link capacities;

$x_{ij}$ Decision variables for link allocation between BTS node i and BSC node j;

$y_{lc}$ Decision variables for choosing the capacity c of E1 (2 Mbps) lines between BSC l and MSC;

$z_{lw}$ Decision variables for BSC l model w choice.

$ct_{ij}$ Link cost between BTS i and BSC j nodes in an analysis time period;

$cm_{lc}$ Link cost of capacity c between BSC l nodes and MSC in an analysis time period;

$cb_w$ BSC model w acquisition cost, considering an analysis time period;

$a_i$ BTS i traffic demand in Erlangs;

$f_c$ Link capacity c in Erlangs;

$e_w$ BSC model w traffic capacity in Erlangs.

*A. Objective Function*

The objective function eq. (1) minimizes total cost of links between BTS and BSC, plus cost of E1 lines between BSC nodes and MSC, plus total cost of BSC's acquisition.

$$minimize \sum_{i \in T} \sum_{j \in B} ct_{ij} x_{ij} + \sum_{l \in B} \sum_{c \in C} cm_{lc} y_{lc} + \sum_{d \in B} \sum_{k \in W} cb_k z_{dk} \quad (1)$$

*B. Restrictions*

In eq. (2), each BTS must be connected to one and only one BSC:

$$\sum_{j \in B} x_{ij} = 1, \forall i \in T \quad (2)$$

In eq. (3), the $y_{lc}$ dimensioning is made. It allows all traffic from BTS assigned to one BSC to flow over its links:

$$\sum_{i \in T} x_{il} a_i \leq \sum_{c \in C} f_c y_{lc}, \forall l \in B \quad (3)$$

In eq. (4), the BSC dimensioning is made accordingly to the given models and the total traffic demand.

$$\sum_{i \in T} x_{ij} a_i \leq \sum_{k \in W} e_k z_{jk}, \forall j \in B \quad (4)$$

$$x_{ij} \in \{0,1\}, \forall i \in T \; \forall j \in B \quad (5)$$

$$y_{lc} \in \{0,1\}, \forall l \in B \; \forall c \in C \quad (6)$$

$$z_{lw} \in \{0,1\}, \forall l \in B \; \forall k \in W \quad (7)$$

## III. MODEL APPLICATION

This model has some issues in real applications that must be observed.

The set of BTS nodes $T$ is known previously because RF engineers make its design as the first step. Its geographical location is determined by coverage and traffic requirements. Its traffic demand can be known previously by measuring other mobile network (old one that is being replaced, or by other overlaid technology such as TDMA (Time Division Multiple Access) or CDMA (Code Division Multiple Access). When such data source is not available, this traffic demands can be estimated by average subscriber traffic and number of subscribers forecast based on population and marketing studies.

The set of BSC nodes $B$ can be generated based on all feasible sites possibilities. The sites that will have a BTS are good candidates, since its space will be already available by rental or buy. Other company buildings can be added to this set. The set $B$ represents all possibilities, and not necessarily the actual BSC allocations. The more options this set $B$ has, the better the allocation of the needed BSC nodes tends to be.

The set $W$ contains the available models of BSC. Normally a BSC manufacturer offers different choices of models. Each one has its capacity in Erlang (as it was modeled here) and price.

The set $C$ is a table of traffic capacities for an integer quantity of E1 lines. Each E1 line has a number of timeslots allocated for voice from the 31 available. Other timeslots are used for signaling and data links. Thus, the first E1 line may have a different number of voice timeslots than the second E1 line, and so on. Each voice timeslot carries 4 compressed voice channels, so called sub-timeslots.

The elements of the set $C$ are calculated by the reverse Erlang B formula, taking the number of voice channels and the defined GoS as incoming data and the traffic as outgoing data. The first element of set $C$ is 0 E1 lines, which lead to 0 Erlang. The second element of set $C$ is 1 E1 line and has a calculated traffic for 4 times the number of timeslots allocated for voice in this E1 line. This is because each timeslot has 4 sub-timeslots. The third element of set $C$ is 2 E1 lines and has the traffic calculated for 4 times the number of timeslots allocated for voice in all 2 E1 lines, and so on. The size of the set $C$ is determined by the maximal capacity of the larger BSC model.

The link costs $ct$ and $cb$ in a given period of analysis must be determined by the transmission network ownership and/or contract. If the transmission network belongs to the own mobile company, its cost can be determined by a set of distance ranges or as a constant times the distance, plus an equipment fixed cost. If the mobile company contracts transmission lines from other company, the costs must be calculated based on specific contractual rules. For instance, discounts based on quantity can be applied.

This integer programming model can be adapted to work with BSC that has maximum number of radio channels capacity, instead of maximum traffic capacity as presented.

### IV. COMPUTATIONAL RESULTS

Simulations were made with many network sizes. The largest network size that could be solved in a reasonable time has about 50 sites. The different generated data caused big differences in the solving time. For instance: The smaller solving time for 50 sites with 3201 integer variables and 150 restrictions was 42.04 seconds, while other equivalent problem instances caused solver to spent more than 30 minutes to solve.

The data was generated using the following assumptions:

- The transmission cost was calculated multiplying the link distance by a constant. Local market cost approximations were used. The cost of $n$ E1 line in the same link is assumed to be $n$ times the cost of each E1 line.

- The BTS and MSC site geographical locations where generated randomly. For each BTS site, a BSC site candidate was generated. The traffic of each BTS was generated randomly from 0 to 80 Erlangs that is the approximated value that a BTS can handle with 1 E1 line.

- The set $C$ was generated with 41 values, from 0 E1 lines until 40 E1 lines. For each capacity, the corresponding traffic was calculated accordingly to the exposed in the model application session (3).

- Three BSC models where used in these simulations: Small, medium and large with 512, 2048 and 4096 Erlangs of capacity respectively. Each one had an acquisition cost compatible to the local market reality.

OPL integrated modeling environment and Cplex 10.0 solver library [9] from Ilog Inc. were used in the simulations. OPL ran in a 64 bits Intel Core 2 Quad processor with 2.4 GHz clock and 4 GB of RAM memory.

Despite the fact that 50 sites is a very small problem instance comparing to hundreds or even thousand sites of the real mobile networks, the simulations shown that this model works properly for the desirable purpose. Varying the costs, more or less BSC were allocated. Each BSC model was correctly chosen accordingly to the total traffic demanded by all BTS allocated to this BSC. The distances were minimized indirectly because of the linear cost by kilometer. The trunk between BSC and MSC was dimensioned to carry the total traffic demand by BSC, and its distance to MSC was effectively considered, since the amount of E1 lines was greater than one.

The 20 problem instances were created and solved for each number of BTS sites varying from 5 until 50 with steps of 5. The data were generated randomly following the premises described in this section. The results are shown in table 1.

## V. SCALABILITY ANALYSIS

Due to the wide range of random generated values, the problem instances have very high complexity variations. Thus, there were problem instances with 40 BTS that could not be solved within a reasonable time threshold. Some times the solver crashed because of memory lack. But, for the same reason, there are problems instances larger than 50 BTS that can be solved in a time interval even smaller than some particular instances of 40 BTS.

The proposed model here is an Integer Programming one. The discrete nature of the variables requires an algorithm like Branch-and-bound, Branch-and-cut or others. This sort of algorithms has an exponential complexity. This fact limits the larger instance size that can be handled. Actual networks often have hundred of BTS that is far beyond the range of this exact method. Aguiar and Pinheiro [13] used Lingo solver library and it was not able to handle problem instances larger than 40 BTS. The adoption of Cplex [9] expanded this boundary to 50 BTS, but it remains too small.

A mean squares non-linear regression of the average times was made to determine the observed asymptotic complexity function. It is shown on eq. 8 and fig. 2.

$$y = 0{,}851 e^{0{,}244x} \qquad (8)$$

The key to break this limitation and turn big network designs feasible is to use approximate approaches. Some methodologies like Lagrangean relaxation in Simple Subgradient, Bundle Methods and Space Dilatation Methods (Shor *et al* [6, 7]) can be used. Rigolon *et al* [3] show that the use of this tool in the first model extends the size of the largest mobile network to be designed. A framework that hybridizes exact methods and meta-heuristics has presented good results in expanding these boundaries in other classes of problems. Nepomuceno, Pinheiro and Coelho [11] used this framework to solve container loading problems. In the same problem category, Pinheiro and Coelho [12] presented a variation of the implementation to work with cutting problems.

| BTS | Var. | Const. | Density | Avg. Time | Std. Deviation |
|---|---|---|---|---|---|
| 5 | 96 | 15 | 9,72% | 50,0 | 12,773 |
| 10 | 241 | 30 | 5,95% | 40,0 | 8,208 |
| 15 | 436 | 45 | 4,43% | 332,0 | 28,802 |
| 20 | 681 | 60 | 3,57% | 853,5 | 86,418 |
| 25 | 976 | 75 | 3,01% | 3561,5 | 371,594 |
| 30 | 1321 | 90 | 2,60% | 19689,0 | 2872,227 |
| 35 | 1716 | 105 | 2,29% | 46287,5 | 4890,274 |
| 40 | 2161 | 120 | 2,05% | 600431,1 | 80263,118 |
| 45 | 2656 | 135 | 1,86% | 363032,5 | 44981,655 |
| 50 | 3201 | 150 | 1,70% | 752724,0 | 87873,235 |

Table 1 - Results

After the model presentation, its application was shown explaining how to relate technical details of the real world with the model's data generation.

In computational results section, size and performance simulations were described. The scalability was analyzed lead to some conclusions. This model by itself can't be used on real networks because of its limitation. Simulation with real networks can't show the optimization potential because small networks can be well designed by human intuition and have smaller costs. Some methodology must be applied to extend the size of the problems to achieve hundred or thousand BTS sites. Thus, the optimization gain can be very effective.

## VII. CONCLUSION

This work gave a solution to the BSS network design problem of mobile GSM carriers capturing its essence in a mathematical model. In introduction section some telecommunications background was given to help understanding the model. Then, the model was presented and explained.

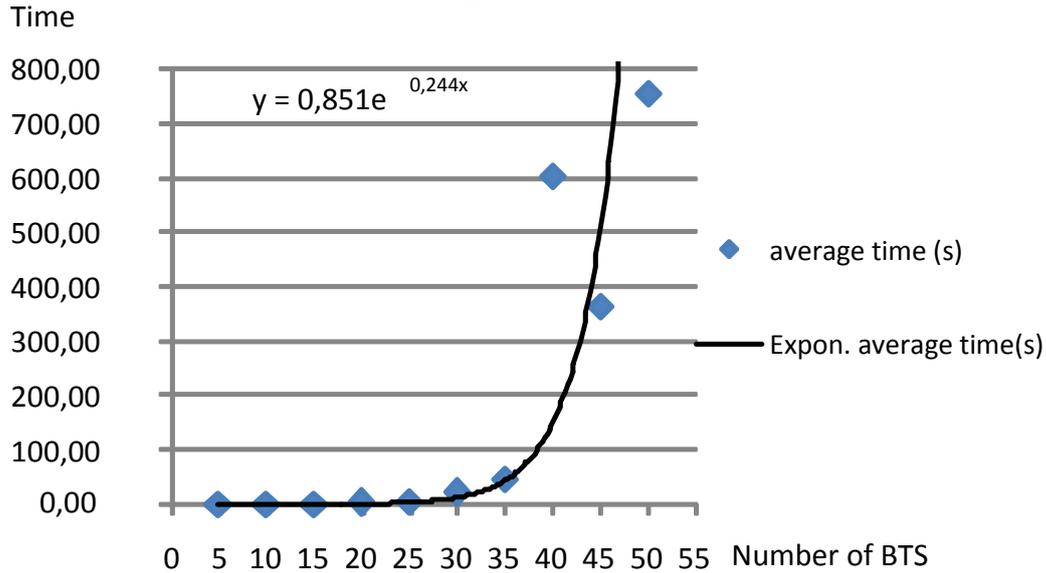

Fig. 2. Average time versus instance size